\documentclass[10pt,prl,aps,twocolumn,superscriptaddress,floatfix,showpacs]{revtex4}

\pdfoutput=1

\usepackage{graphicx}
\usepackage{amssymb}
\usepackage{bm}% bold math
\usepackage{bbold}% bold math

\begin{document}

\title{RVB superconductors with fermionic projected entangled pair states}

\author{Didier Poilblanc}
\affiliation{Laboratoire de Physique Th\'eorique, C.N.R.S. and Universit\'e de Toulouse, 31062 Toulouse, France}

\author{Philippe Corboz}
\affiliation{Theoretische Physik, ETH-Zurich, 8093 Zurich, Switzerland}
\affiliation{Institute for Theoretical Physics, University of Amsterdam,
Science Park 904 Postbus 94485, 1090 GL Amsterdam, The Netherlands}

\author{Norbert \surname{Schuch}}
\affiliation{Institut f\"ur Quanteninformation, RWTH Aachen, D-52056 Aachen, Germany}

\author{J.~Ignacio \surname{Cirac}}
\affiliation{Max-Planck-Institut f{\"{u}}r Quantenoptik,
Hans-Kopfermann-Str.\ 1, D-85748 Garching, Germany}

\date{\today}

\begin{abstract}
  We construct a family of simple fermionic projected entangled pair states (fPEPS) on
  the square lattice with bond dimension $D=3$ which
  are exactly hole-doped resonating valence bond (RVB) wavefunctions
  with short-range singlet bonds. Under doping the insulating RVB spin liquid
  evolves immediately into a superconductor with mixed $d+is$ pairing symmetry whose pair amplitude 
  grows as the square-root of the doping. The relative weight between $s$-wave and $d$-wave components
  can be controlled by a single variational parameter $c$.  
 We optimize our ansatz w.r.t. $c$ for the frustrated $t-J_1-J_2$ model (including both nearest and next-nearest neighbor antiferromagnetic interactions $J_1$ and $J_2$, respectively) for $J_2\simeq J_1/2$ and obtain an energy 
very close to the infinite-PEPS
 state (using full update optimization and same bond dimension).
 The orbital symmetry of the optimized RVB superconductor has 
 predominant $d$-wave character, although we argue a residual (complex $s$-wave) time reversal symmetry breaking component 
 should always be present.
Connections of the results to the physics of superconducting cuprates and pnictides are outlined. 
 
\end{abstract}
\pacs{75.10.Kt,75.10.Jm}
\maketitle

%\section*{Introduction}
{\bf Introduction} -- The concept of resonant valence bond (RVB) state
was first introduced by Anderson~\cite{RVB0} to describe a possible 
quantum disordered ground
state in the triangular lattice $S=1/2$ Heisenberg antiferromagnet.
Insulating RVB states are commonly defined as equal weight superpositions of hardcore coverings
of nearest-neighbor (NN) or short-range singlets. Such RVB states are
now well understood thanks to a large variety of large-scale approaches as improved Monte Carlo sampling
scheme for valence bonds, use of representations in terms of simple Projected Entangled Pair States (PEPS)
or mapping into projected BCS wave functions~: while the RVB state on the square lattice 
is critical~\cite{albuquerque,tang,wang} 
their analogs on kagome and triangular lattices were shown~\cite{norbert1,didier,wildeboer,yangfan} 
to be $\mathbb{Z}_2$ gapped spin liquids.
Interestingly, recent numerical results pointed towards an algebraic spin liquid 
in the frustrated spin 1/2 $J_1-J_2$ antiferromagnetic (AFM) Heisenberg model
on the square lattice~\cite{wang},
% for a ratio of the next-NN to the NN exchange couplings $J_2/J_1\sim 0.5$,
which might also be the sign of a critical point between a N\'eel AFM state and a dimer state~\cite{anders_j1j2}. 
%~\cite{jiang,wang}.

\begin{figure}
\begin{center}
\includegraphics[width=6.5cm,angle=0]{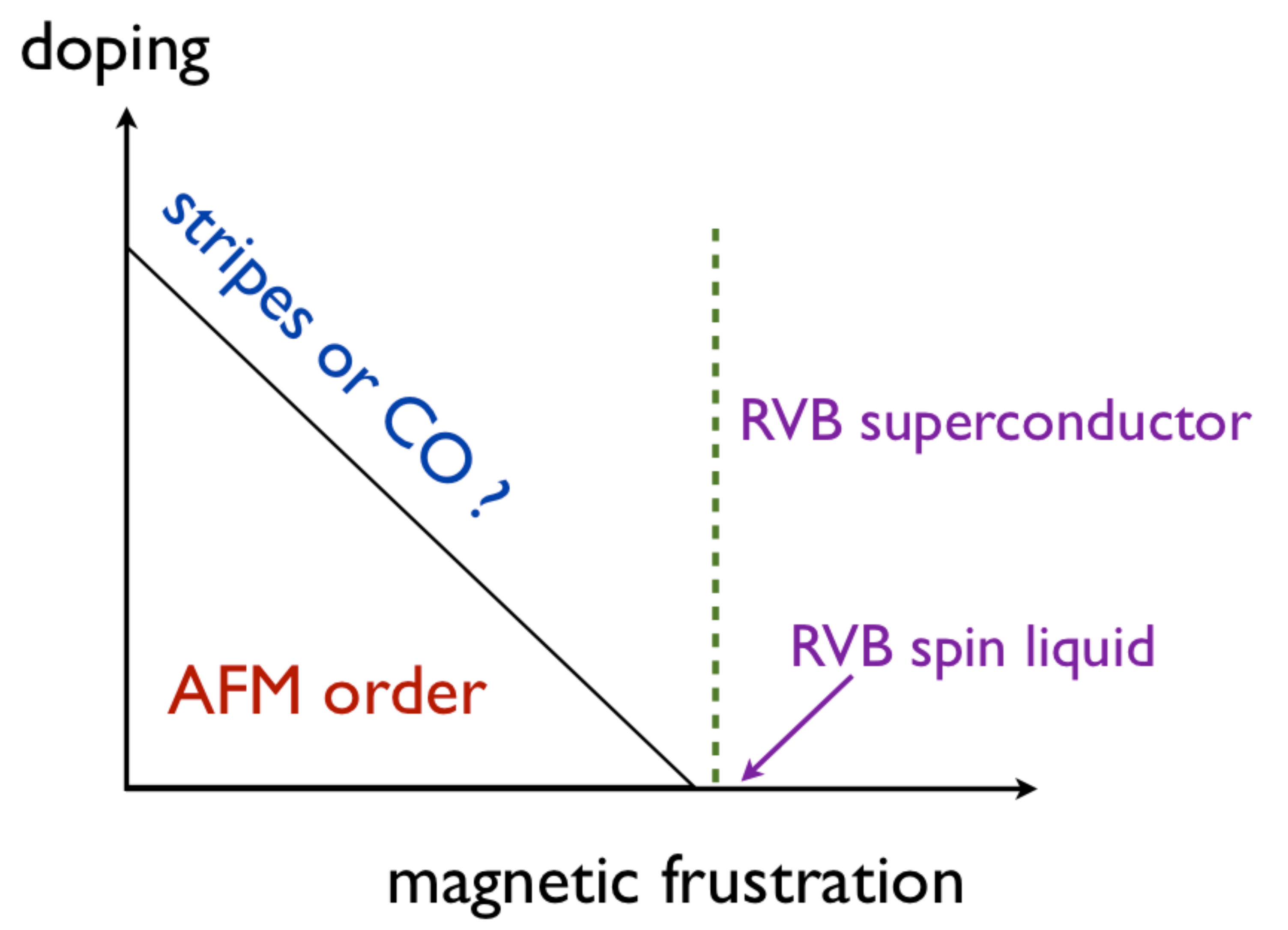}
\caption{Schematic ($T=0$) phase diagram of an AFM Mott insulator as a function of doping 
and magnetic frustration. Doping the AFM phase might involve complex 
phases such as ``stripes" or charge ordered (CO) states. If a RVB liquid is stabilized by magnetic frustration
it would evolve naturally under doping (dotted line) into a RVB superconductor.
  }
\label{Fig:PD}
\end{center}
\vskip -0.5cm
\end{figure}

The RVB ideas have naturally been extended away from the 1/2-filled Mott insulator at
electron density $n<1$ and a ``RVB superconductor"
has been proposed as a simple mechanism in the high-temperature cuprates superconductors~\cite{RVB,patrick}. 
However, neutron scattering~\cite{neutrons}, scanning tunneling microscopy~\cite{stm}, nuclear magnetic resonance~\cite{nmr}
and resonant soft X-ray scattering~\cite{xray} experiments have shown that the cuprate parent AFM phase (generically) 
evolves under doping into 
complex intertwined phases involving e.g. charge-stripe or nematic orders.
% as originally predicted by mean-field calculations of Hubbard models~\cite{zaanen89,poilblanc89}. 
As suggested in Fig.~\ref{Fig:PD} a (quantum disordered) 
RVB spin liquid stabilized by magnetic frustration would however have a more straightforward evolution under doping.
Nevertheless, little is known about RVB states in which charge fermionic degrees of freedom come into play, 
although preliminary work has been done on (fermionic) doped dimer liquids lacking
spin-$SU(2)$ symmetry~\cite{qdm1,qdm2}. In this paper, we introduce a general {\it fermionic} doped RVB state written as a $D=3$
Projected Entangled Paired State (PEPS) on the square lattice. This state is obtained by (i) rewriting the nearest-neighbor 
RVB PEPS~\cite{norbert1,didier,wang} {\it in the fermion representation}, (ii) introducing vacant sites corresponding to
a finite (average) density $x=1-n$ of doped holes (or ``holons") and, finally, (iii) introducing longer range singlets
(e.g. along diagonal bonds) {\it next} to some of the doped holes in a way that meets all lattice symmetries. 
Step (iii) is controlled by a single variational
parameter $c$. We show that this state is a superconductor (hence breaking charge U(1) symmetry) which 
(generically) inherits the 
mixed $d+is$ orbital symmetry of its parent (insulating) spin liquid, 
although its s-wave component  is severely suppressed by (iii). Optimizing the hole kinetic energy
w.r.t. $c$ we obtain a good ansatz for the fermionic hole-doped frustrated spin-1/2 $J_1-J_2$ AFM
Heisenberg model at $J_2=0.5J_1$ and $J_1=0.4t$ (so-called $t-J_1-J_2$ model).
The idea is therefore to introduce a simple yet competing
wavefunction that enables us to understand
unconventional superconducting ground states of strongly correlated fermions.

\begin{figure}
\begin{center}
\includegraphics[width=6.5cm,angle=0]{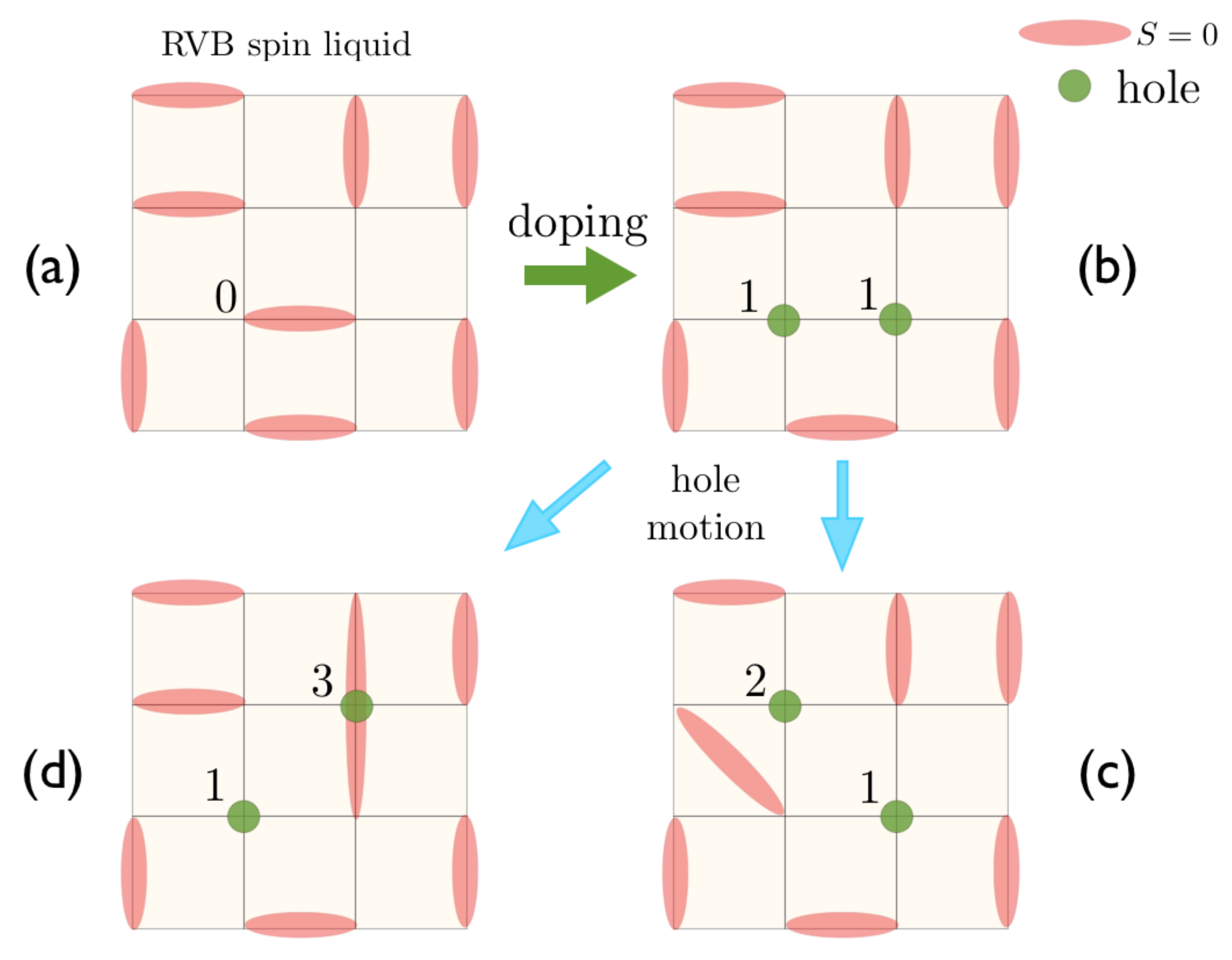}
\caption{ (a-d) Red segments represent
singlet pairs of {\it physical} spin-$1/2$s. Green dots represent doped holes which can hop
in the $x$- or $y$-directions. 
In the RVB insulator (a) or superconductor (b-d), {\it resonances} between many such 
configurations occur.
fPEPS tensor elements encode each of the site configurations labeled from $0$ to $3$. 
  }
\label{Fig:RVB}
\end{center}
\vskip -0.5cm
\end{figure}

{\bf Doped RVB states in PEPS formalism} -- In the PEPS with bond dimension
$D=3$ we consider, each physical site has 4 virtual spins attached,
each of which spans a virtual dimension of spin $1/2\oplus 0$. On every bond, 
every pair of the NN virtual spins is projected to a virtual spin singlet state,
$
|\mathcal{S}\rangle = |01\rangle-|10\rangle +|22\rangle,
$
where the virtual indices ``0,1'' span the subspace of spin $1/2$ and
virtual index ``2'' spans the subspace of spin 0~\cite{note1}. 
At each site, one considers a projector $\mathcal{P}$ to enforce 
the local physical degrees of freedom. Finally, contracting the virtual index of each $\mathcal{S}$ at the bond 
and each $\mathcal{P}$ at the vertex, yields the desired PEPS state.

We start with $\mathcal{P}=\mathcal{P}_0$ which maps one of the virtual spin
$1/2$ subspace onto the physical spin $1/2$ state (leaving the
rest of virtual spins in the ``2'' state) and gives exactly the equal
weight NN RVB state of Fig.~\ref{Fig:RVB}(a)~\cite{norbert1,didier}.
Next, a fermionic character is assigned to both virtual and physical spins $1/2$ following the 
procedure of Ref.~\onlinecite{corboz} to construct a fermionic PEPS (fPEPS). To recover the previous bosonic RVB state 
one must now include an extra $i$ complex factor on (let say) the vertical singlets, i.e. 
$|\mathcal{S}\rangle = i(|01\rangle-|10\rangle) +|22\rangle$,
leading to $d+is$ point group (or orbital) symmetry~\cite{kotliar}.
One can now dope the fermonic $d+is$ RVB insulator by simply
enlarging the physical space to vacant sites labeled e.g. by the index ``2".
Adding a projector $\mathcal{P}_1$ enforcing all (both virtual and physical) spins in the ``2'' state
generates local hole configurations labelled by ``1" in Figs.~\ref{Fig:RVB}(b-d). 
However, the doped RVB state characterized only by $\mathcal{P}=\mathcal{P}_0+\lambda\mathcal{P}_1$ 
(where $\lambda$ plays the role of a ``chemical potential") has zero expectation value of 
the hole kinetic operator between NN sites,
 \begin{equation}
\label{projector}
H_K=-t \sum_{\big<ij\big>} {P}_G \, c_{i,\sigma}^\dagger c_{j,\sigma} {P}_G + H.C.
\end{equation}
where $c_{i,\sigma}^\dagger$ creates a fermion (electron) of spin $\sigma$ at site $i$ 
and ${P}_G$ is the Gutzwiller projector enforcing the physical Hilbert space that excludes doubly occupied sites,
and therefore can not be a physical representation of doped Mott insulators.  
To circumvent such a problem we allow singlet pairings between sites on the {\it same sublattice} 
in the immediate vicinity of some holes, as e.g. holes ``2" and ``3"  
in Fig.~\ref{Fig:RVB}(c) and Fig.~\ref{Fig:RVB}(d), respectively.
This can be achieved by extending the idea of Ref.~\onlinecite{wang} to the doped case and use 
a "teleportation" projector,
\begin{equation}
\nonumber
  \mathcal{P}_2=\sum_{i\neq j\neq k\neq l}|2\rangle\langle 2 2|_{ij}\otimes\langle\epsilon|_{kl},
\end{equation}
where $|\epsilon\rangle_{kl}$
is a singlet between sites $k$ and $l$ at distance $\sqrt{2}$ (choosing an anticlockwise orientation around the hole)
or $2$ (oriented from left to right, bottom to top).
A general doped RVB wavefunction is a parameter $c$ weighted 
combination of projectors
$\mathcal{P}\equiv\mathcal{P}_0+\lambda(\mathcal{P}_1+c\mathcal{P}_2)$ at each vertex
($\lambda\in\mathbb{C}$, $c\in\mathbb{R}$) traced out with the bond singlets $\mathcal{S}$ at
each bond and including the fermion signs according to Ref.~\onlinecite{corboz}.
Note that this RVB wavefunction is defined in the {\it grand-canonical} ensemble with 
fluctuating particle number.  

\begin{figure}
\begin{center}
\includegraphics[width=8.5cm,angle=0]{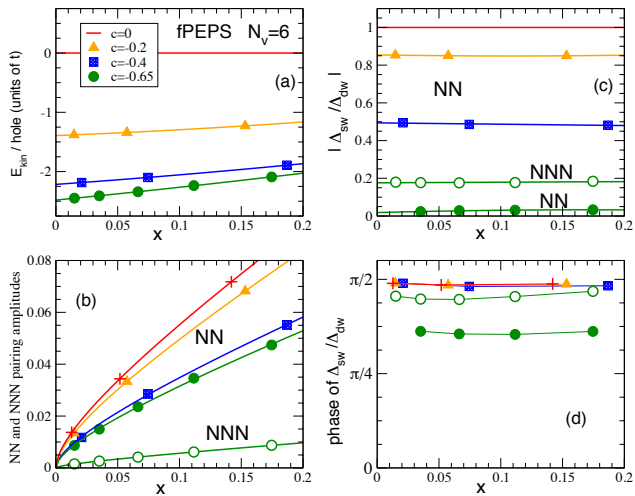}
\caption{Hole kinetic energy (a) and NN pairings (b-d) of the RVB superconductor vs doping,
for several values of the parameter $c$, computed on an infinite $N_v=6$ cylinder. For $c=-0.65$, data for the (s-wave) NNN pairing are also shown (open symbols).
Ratio of the amplitudes (c) and relative phase (d) between
the $s$-wave and the NN $d$-wave components.
$c=0$ is a NN pure $d+is$ superconductor with zero kinetic energy (red lines).  }
\label{Fig:kin-pairing}
\end{center}
\vskip -0.5cm
\end{figure}

{\bf Superconducting order} --
Using $\lambda$ to tune the hole density $x$, we have investigated the properties of this one-parameter RVB family 
on an infinitely long cylinder with a circumference of $N_v=6$ unit cells.
As soon as $c\ne 0$, $\big< H_K\big>\ne 0$ as shown in Fig.~\ref{Fig:kin-pairing}(a). A kinetic energy per hole 
$\big< H_K\big> / x$ as large
as $\sim -2.5 t$ can be obtained for $c\simeq -0.65$ (when $x\rightarrow 0$). 

By construction the doped RVB ansatz breaks the charge U(1) symmetry. This is reflected by finite superconducting (SC)
(singlet) pairings $\Delta_{ij}\equiv\big<c_{i,\uparrow}c_{j,\downarrow}\big>$ for any finite doping $x$.
Changing the phase of $\lambda$ changes the (global) phase of the SC order parameter, i.e. 
$\lambda\rightarrow \exp{(i\phi)}\lambda$ leads to $\Delta_{ij}\rightarrow\exp{(2i\phi)}\Delta_{ij}$.
The SC amplitude $|\Delta_{ij}|$ is largest when $i$ and $j$ are NN sites. Note that the NN pairings $\Delta_{X}$
and $\Delta_{Y}$ along the horizontal and vertical directions, respectively, differ slightly because of the
finite cylinder circumference.  Their average amplitude $\frac{1}{2}(|\Delta_{X}|+|\Delta_{Y}|)$ plotted in Fig.~\ref{Fig:kin-pairing}(b) shows that $\Delta_{ij}$ grows like $\sqrt{x}$.

We now examine the orbital symmetry of the NN pairing field that we decompose into its (a priori complex)
$s$- and $d$-wave components, $\Delta_{\rm sw}=\frac{1}{2}(\Delta_X+\Delta_Y)$
and $\Delta_{\rm dw}=\frac{1}{2}(\Delta_X-\Delta_Y)$, respectively.
We define the ratio $R=\Delta_{\rm sw}/\Delta_{\rm dw}=|R|\exp{(i\Psi)}$ and plot its amplitude $|R|$ 
and phase $\Psi$ in Fig.~\ref{Fig:kin-pairing}(c) and Fig.~\ref{Fig:kin-pairing}(d), respectively.
Note that only the {\it relative} phase $\Psi$ between the $\Delta_{\rm sw}$ and $\Delta_{\rm dw}$ components
matters since the overall
SC phase can be changed by changing the phase of $\lambda$. 
First, we  find that $\Psi$ is almost exactly $\pi/2$ -- as expected
for $d+is$ orbital symmetry.
For $c=0$ the RVB superconductor is a pure NN $d+is$ superconductor with exactly the same $s$- and $d$-wave amplitudes. 
The $s$-wave component is suppressed by increasing the parameter $c$, as the single hole kinetic energy increases (in magnitude).
Interestingly, weak pairing between next-nearest neighbor (NNN) sites 
(i.e. along the plaquette diagonals) develops for increasing $c$,
as shown in Fig.~\ref{Fig:kin-pairing}(b) and \ref{Fig:kin-pairing}(c). The NNN pairing has $s$-wave orbital symmetry and,
as the NN $s$-wave component, a relative phase of $\sim \pi/2$ w.r.t. the leading NN $d$-wave component, as shown in Fig.~\ref{Fig:kin-pairing}(d).
The superconducting coherence length which is, strictly speaking, of one lattice spacing for $c=0$ (the NNN pairing vanishes) grows 
for increasing $c$, yielding to a strong decrease of the NN Coulomb repulsion~\cite{supplement}  (together with a gain of kinetic energy). 

\begin{figure}
\begin{center}
\includegraphics[width=8.5cm,angle=0]{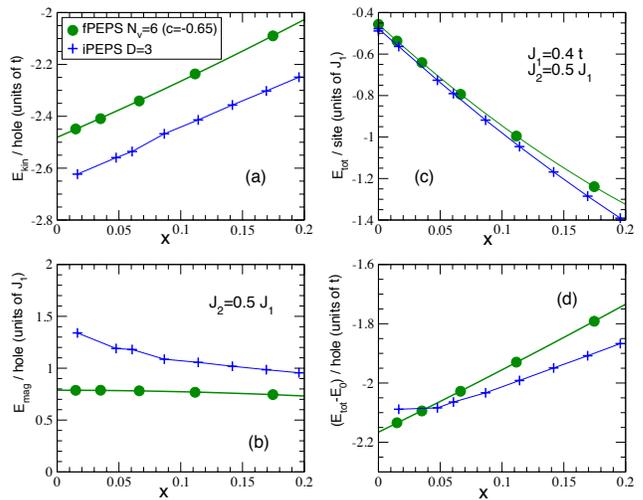}
\caption{ (a-d) Variational energies of the RVB superconductor (with $c=-0.65$) as a function of doping $x$.
(a) kinetic energy per hole; (b) magnetic energy cost per hole; total energy per site (c) and per hole (subtracting the $x=0$ contribution) (d). 
Comparisons with the $D=3$ iPEPS state optimized using the full update scheme (crosses) are shown.}
\label{Fig:energy}
\end{center}
\vskip -0.5cm
\end{figure}

{\bf Optimized RVB superconductor} -- With the schematic phase diagram of 
Fig.~\ref{Fig:PD} in mind, we now introduce the ``frustrated $t-J$ model",
\begin{equation}
\label{projector}
H=H_K+J_1\sum_{\big<ij\big>} {\bf S}_i\cdot{\bf S}_j + J_2\sum_{\big<\big<kl\big>\big>} {\bf S}_k\cdot{\bf S}_l
\label{Eq:tJ}
\end{equation}
involving AFM interactions $J_1$ and $J_2$ between NN and NNN sites, respectively. 
For $J_2=0$, recent infinite-PEPS (iPEPS) calculations~\cite{corboz2} revealed an extremely close competition between a uniform $d$-wave superconducting state and different stripe states under doping the N\'eel AFM insulator. Here, we instead fix $J_2=0.5J_1$
for which we expect the half-filled GS to be a critical spin liquid well approximated by the simple NN 
RVB~\cite{wang} discussed above.  Therefore, at finite hole density, the doped (fermionic) \hbox{$d+is$} RVB state 
naturally becomes a promising variational candidate for Hamiltonian (\ref{Eq:tJ}).
We find that the amplitude $|\big<H_K\big>|$ in this state of the kinetic energy is maximized for $c\simeq -0.65$,
and the corresponding data are shown in Fig.~\ref{Fig:energy}(a). A crude fit gives that the kinetic energy per hole
behaves as $e_k\simeq (-2.5+2.3 x)t$. On the other hand, mobile holes, by perturbing the spin RVB background, cost magnetic energy
that is balancing the gain of kinetic energy.  
The magnetic energy cost (per hole) can be quantitatively defined as $e_m=(|\big<H_m\big>_x|-|\big<H_m\big>_0|)/x$ where 
$H_m=H-H_K$ is the magnetic part of (\ref{Eq:tJ}) and $\big<\cdots\big>_x$ is the expectation value in the RVB state at
(average) doping $x$. A crude fit based on Fig.~\ref{Fig:energy}(b) gives $e_m\simeq (0.8-0.3x)J_1$
for $J_2=0.5 J_1$. The overall total energy $\big<H_m\big>_0/N_s +x(e_k+e_m)$, normalized by the number of sites $N_s$, is shown in 
Fig.~\ref{Fig:energy}(c) and the hole contribution $e_k+e_m$ in Fig.~\ref{Fig:energy}(d), assuming $J_1=0.4 t$.

We have compared in Figs.~\ref{Fig:energy}(a-d) the results for the constructed RVB state to iPEPS calculations~\cite{supplement} 
using a 4 site unit cell, a ``full'' update scheme and the same bond dimension 
$D=3$. As seen in Fig.~\ref{Fig:energy}(c), the total energy of the $d+is$ RVB is very close to
the optimized iPEPS energy (starting from a random configuration). In fact, using the local tensor of the RVB state
as a starting point, only minor improvement is obtained with the iPEPS optimization scheme. The $d+is$ RVB
state is therefore a very good variational ansatz.

{\bf Discussions} -- Although magnetic frustration is weak in the high-temperature 
cuprate superconductors~\cite{coldea,malrieu}, it certainly plays an important role in the iron pnictides superconductors
for which a frustrated Heisenberg model has been proposed~\cite{yao,abrahams}, despite the multi-orbital 
character of the materials. 
In any case, the uniform RVB superconductor we propose here might extend to a larger region of the
phase diagram of Fig.~\ref{Fig:PD} than just the dotted line (corresponding to the 
present work) and might have some relevance to the physics of high-temperature superconducting materials. 
At finite temperature, superconducting phase coherence may be lost while singlet pairing still occurs, 
hence providing a simple picture for the ``pseudo-gap phase" of the cuprates. 
Interestingly, our scenario predicts a small residual {\it imaginary} $s$-wave component 
of the pairing field (at zero-temperature) which would give rise to (very weak) broken time reversal 
symmetry (BTRS). Note that early claims of BTRS were made in the cuprates based on angular 
resolved photoemission (ARPES) experiments~\cite{campuzano} but recent polar Kerr-effect
measurements~\cite{kapitulnik}  
have been interpreted in terms of chiral charge ordering.
Incidentely, no orbital currents  
-- as suggested by polarized neutron diffraction experiments~\cite{fauque} -- occurs in our RVB superconductor.
 
{\bf Acknowledgment} --  D.P. acknowledges the
 NQPTP ANR-0406-01 grant (French Research Council) for support.
The computational results presented have been achieved 
using the CALMIP Hyperion Cluster (Toulouse). N.S. acknowledges support from the
Alexander von Humboldt foundation. I.C. acknowledges support from the EU project SIQS.


\begin{thebibliography}{7}%
\makeatletter
\providecommand \@ifxundefined [1]{%
 \@ifx{#1\undefined}
}%
\providecommand \@ifnum [1]{%
 \ifnum #1\expandafter \@firstoftwo
 \else \expandafter \@secondoftwo
 \fi
}%
\providecommand \@ifx [1]{%
 \ifx #1\expandafter \@firstoftwo
 \else \expandafter \@secondoftwo
 \fi
}%
\providecommand \natexlab [1]{#1}%
\providecommand \enquote  [1]{``#1''}%
\providecommand \bibnamefont  [1]{#1}%
\providecommand \bibfnamefont [1]{#1}%
\providecommand \citenamefont [1]{#1}%
\providecommand \href@noop [0]{\@secondoftwo}%
\providecommand \href [0]{\begingroup \@sanitize@url \@href}%
\providecommand \@href[1]{\@@startlink{#1}\@@href}%
\providecommand \@@href[1]{\endgroup#1\@@endlink}%
\providecommand \@sanitize@url [0]{\catcode `\\12\catcode `\$12\catcode
  `\&12\catcode `\#12\catcode `\^12\catcode `\_12\catcode `\%12\relax}%
\providecommand \@@startlink[1]{}%
\providecommand \@@endlink[0]{}%
\providecommand \url  [0]{\begingroup\@sanitize@url \@url }%
\providecommand \@url [1]{\endgroup\@href {#1}{\urlprefix }}%
\providecommand \urlprefix  [0]{URL }%
\providecommand \Eprint [0]{\href }%
\providecommand \doibase [0]{http://dx.doi.org/}%
\providecommand \selectlanguage [0]{\@gobble}%
\providecommand \bibinfo  [0]{\@secondoftwo}%
\providecommand \bibfield  [0]{\@secondoftwo}%
\providecommand \translation [1]{[#1]}%
\providecommand \BibitemOpen [0]{}%
\providecommand \bibitemStop [0]{}%
\providecommand \bibitemNoStop [0]{.\EOS\space}%
\providecommand \EOS [0]{\spacefactor3000\relax}%
\providecommand \BibitemShut  [1]{\csname bibitem#1\endcsname}%
\let\auto@bib@innerbib\@empty
%</preamble>
\bibitem [{\citenamefont {Calzado}\ and\ \citenamefont
  {Malrieu}(2001)}]{malrieu2001}%
  \BibitemOpen
  \bibfield  {author} {\bibinfo {author} {\bibfnamefont {C.~J.}\ \bibnamefont
  {Calzado}}\ and\ \bibinfo {author} {\bibfnamefont {J.-P.}\ \bibnamefont
  {Malrieu}},\ }\href {http://link.aps.org/doi/10.1103/PhysRevB.63.214520}
  {\bibfield  {journal} {\bibinfo  {journal} {Phys. Rev. B}\ }\textbf {\bibinfo
  {volume} {63}},\ \bibinfo {pages} {214520} (\bibinfo {year}
  {2001})}\BibitemShut {NoStop}%
\bibitem [{\citenamefont {Poilblanc}(1994)}]{poilblanc1994}%
  \BibitemOpen
  \bibfield  {author} {\bibinfo {author} {\bibfnamefont {D.}~\bibnamefont
  {Poilblanc}},\ }\href {http://link.aps.org/doi/10.1103/PhysRevB.49.1477}
  {\bibfield  {journal} {\bibinfo  {journal} {Phys. Rev. B}\ }\textbf {\bibinfo
  {volume} {49}},\ \bibinfo {pages} {1477} (\bibinfo {year}
  {1994})}\BibitemShut {NoStop}%
\bibitem [{\citenamefont {Corboz}\ \emph
  {et~al.}(2010{\natexlab{a}})\citenamefont {Corboz}, \citenamefont {Orus},
  \citenamefont {Bauer},\ and\ \citenamefont {Vidal}}]{corboz2010}%
  \BibitemOpen
  \bibfield  {author} {\bibinfo {author} {\bibfnamefont {P.}~\bibnamefont
  {Corboz}}, \bibinfo {author} {\bibfnamefont {R.}~\bibnamefont {Orus}},
  \bibinfo {author} {\bibfnamefont {B.}~\bibnamefont {Bauer}}, \ and\ \bibinfo
  {author} {\bibfnamefont {G.}~\bibnamefont {Vidal}},\ }\href {\doibase
  10.1103/PhysRevB.81.165104} {\bibfield  {journal} {\bibinfo  {journal} {Phys.
  Rev. B}\ }\textbf {\bibinfo {volume} {81}},\ \bibinfo {pages} {165104}
  (\bibinfo {year} {2010}{\natexlab{a}})}\BibitemShut {NoStop}%
\bibitem [{\citenamefont {Corboz}\ \emph
  {et~al.}(2010{\natexlab{b}})\citenamefont {Corboz}, \citenamefont {Jordan},\
  and\ \citenamefont {Vidal}}]{corboz2010-nn}%
  \BibitemOpen
  \bibfield  {author} {\bibinfo {author} {\bibfnamefont {P.}~\bibnamefont
  {Corboz}}, \bibinfo {author} {\bibfnamefont {J.}~\bibnamefont {Jordan}}, \
  and\ \bibinfo {author} {\bibfnamefont {G.}~\bibnamefont {Vidal}},\ }\href
  {\doibase 10.1103/PhysRevB.82.245119} {\bibfield  {journal} {\bibinfo
  {journal} {Phys. Rev. B}\ }\textbf {\bibinfo {volume} {82}},\ \bibinfo
  {pages} {245119} (\bibinfo {year} {2010}{\natexlab{b}})}\BibitemShut
  {NoStop}%
\bibitem [{\citenamefont {Corboz}\ and\ \citenamefont
  {Mila}(2013)}]{Corboz13_shastry}%
  \BibitemOpen
  \bibfield  {author} {\bibinfo {author} {\bibfnamefont {P.}~\bibnamefont
  {Corboz}}\ and\ \bibinfo {author} {\bibfnamefont {F.}~\bibnamefont {Mila}},\
  }\href {\doibase 10.1103/PhysRevB.87.115144} {\bibfield  {journal} {\bibinfo
  {journal} {Phys. Rev. B}\ }\textbf {\bibinfo {volume} {87}},\ \bibinfo
  {pages} {115144} (\bibinfo {year} {2013})}\BibitemShut {NoStop}%
\bibitem [{\citenamefont {Wang}\ \emph {et~al.}(2013)\citenamefont {Wang},
  \citenamefont {Poilblanc}, \citenamefont {Gu}, \citenamefont {Wen},\ and\
  \citenamefont {Verstraete}}]{wang13}%
  \BibitemOpen
  \bibfield  {author} {\bibinfo {author} {\bibfnamefont {L.}~\bibnamefont
  {Wang}}, \bibinfo {author} {\bibfnamefont {D.}~\bibnamefont {Poilblanc}},
  \bibinfo {author} {\bibfnamefont {Z.-C.}\ \bibnamefont {Gu}}, \bibinfo
  {author} {\bibfnamefont {X.-G.}\ \bibnamefont {Wen}}, \ and\ \bibinfo
  {author} {\bibfnamefont {F.}~\bibnamefont {Verstraete}},\ }\href
  {http://link.aps.org/doi/10.1103/PhysRevLett.111.037202} {\bibfield
  {journal} {\bibinfo  {journal} {{Phys. Rev. Lett.}}\ }\textbf {\bibinfo
  {volume} {111}},\ \bibinfo {pages} {037202} (\bibinfo {year}
  {2013})}\BibitemShut {NoStop}%
\bibitem [{\citenamefont {Corboz}\ \emph {et~al.}(2014)\citenamefont {Corboz},
  \citenamefont {Rice},\ and\ \citenamefont {Troyer}}]{corboz14}%
  \BibitemOpen
  \bibfield  {author} {\bibinfo {author} {\bibfnamefont {P.}~\bibnamefont
  {Corboz}}, \bibinfo {author} {\bibfnamefont {T.~M.}\ \bibnamefont {Rice}}, \
  and\ \bibinfo {author} {\bibfnamefont {M.}~\bibnamefont {Troyer}},\ }\href
  {http://arxiv.org/abs/1402.2859} {\bibfield  {journal} {\bibinfo  {journal}
  {{arXiv:1402.2859} [cond-mat]}\ } (\bibinfo {year} {2014})}\BibitemShut
  {NoStop}%
\end{thebibliography}%


\begin{thebibliography}{99}
\expandafter\ifx\csname
natexlab\endcsname\relax\def\natexlab#1{#1}\fi
\expandafter\ifx\csname bibnamefont\endcsname\relax
  \def\bibnamefont#1{#1}\fi
\expandafter\ifx\csname bibfnamefont\endcsname\relax
  \def\bibfnamefont#1{#1}\fi
\expandafter\ifx\csname citenamefont\endcsname\relax
  \def\citenamefont#1{#1}\fi
\expandafter\ifx\csname url\endcsname\relax
  \def\url#1{\texttt{#1}}\fi
\expandafter\ifx\csname urlprefix\endcsname\relax\def\urlprefix{URL
}\fi \providecommand{\bibinfo}[2]{#2}
\providecommand{\eprint}[2][]{\url{#2}}

\bibitem{RVB0}
  \bibinfo{author}{\bibfnamefont{P.~W.}~\bibnamefont{Anderson}},
  \bibinfo{title}{\bibfnamefont{Resonating valence bonds: A new kind of insulator?}},
  \bibinfo{journal}{Mat. Res. Bull.} \textbf{\bibinfo{volume}{8}},
  \bibinfo{pages}{153} (\bibinfo{year}{1973}).

% \bibitem{liang}
 % \bibinfo{author}{\bibfnamefont{S.~Liang}},
  %\bibinfo{author}{\bibfnamefont{B.~Doucot}} \bibfnamefont{and}
 % \bibinfo{author}{\bibfnamefont{P.~W.~Anderson}},
 % \bibinfo{title}{\bibfnamefont{Some New Variational Resonating-Valence-Bond-Type Wave Functions for the Spin-$\frac{1}{2}$ Antiferromagnetic Heisenberg Model on Square Lattice}},
  %\bibinfo{journal}{Phys.~Rev.~Lett.} \textbf{\bibinfo{volume}{61}},
  %\bibinfo{pages}{365} (\bibinfo{year}{1988}).

\bibitem{albuquerque}
  \bibinfo{author}{\bibfnamefont{A.~F.~Albuquerque}} \bibfnamefont{and}
  \bibinfo{author}{\bibfnamefont{F.~Alet}},
  \bibinfo{title}{\bibfnamefont{Critical correlations for short-range valence-bond wavefunctions on the square lattice}},
  \bibinfo{journal}{Phys.~Rev.~B} \textbf{\bibinfo{volume}{82}},
  \bibinfo{pages}{180408R} (\bibinfo{year}{2010}).

 \bibitem{tang}
  \bibinfo{author}{\bibfnamefont{Y.~Tang}},
  \bibinfo{author}{\bibfnamefont{A.~W.~Sandvik}} \bibfnamefont{and}
  \bibinfo{author}{\bibfnamefont{C.~L.~Henley}},
  \bibinfo{title}{\bibfnamefont{Properties of resonating valence bond spin liquids and critical dimer models}},
  \bibinfo{journal}{Phys.~Rev.~B} \textbf{\bibinfo{volume}{84}},
  \bibinfo{pages}{174427} (\bibinfo{year}{2011}).

\bibitem{wang}
  \bibinfo{author}{\bibfnamefont{Ling~Wang}},
  \bibinfo{author}{\bibfnamefont{Didier~Poilblanc}},
  \bibinfo{author}{\bibfnamefont{Zheng-Cheng Gu}} 
\bibinfo{author}{\bibfnamefont{Xiao-Gang Wen}} 
  \bibfnamefont{and}
  \bibinfo{author}{\bibfnamefont{Frank Verstraete}},
  \bibinfo{title}{\bibfnamefont{Constructing gapless spin liquid state for the spin-1/2 J1-J2 Heisenberg model on a square lattice }},
  \bibinfo{journal}{Phys.~Rev.~Lett.} \textbf{\bibinfo{volume}{111}},
  \bibinfo{pages}{037202} (\bibinfo{year}{2013}).

\bibitem{norbert1}
  \bibinfo{author}{\bibfnamefont{N.~Schuch}},
  \bibinfo{author}{\bibfnamefont{D.~Poilblanc}},
  \bibinfo{author}{\bibfnamefont{J.~I.~Cirac}} \bibfnamefont{and}
  \bibinfo{author}{\bibfnamefont{D.~Perez-Garcia}},
  \bibinfo{title}{\bibfnamefont{Resonating valence bond states in the PEPS formalism}},
  \bibinfo{journal}{Phys.~Rev.~B} \textbf{\bibinfo{volume}{86}},
  \bibinfo{pages}{115108} (\bibinfo{year}{2012}).


\bibitem{didier}
  \bibinfo{author}{\bibfnamefont{D.~Poilblanc}},
  \bibinfo{author}{\bibfnamefont{N.~Schuch}},
  \bibinfo{author}{\bibfnamefont{D.~Perez-Garcia}} \bibfnamefont{and}
  \bibinfo{author}{\bibfnamefont{J.~I.~Cirac}},
  \bibinfo{title}{\bibfnamefont{Topological and entanglement
      properties of resonating valence bond wave functions}},
  \bibinfo{journal}{Phys.~Rev.~B} \textbf{\bibinfo{volume}{86}},
  \bibinfo{pages}{014404} (\bibinfo{year}{2012}).

\bibitem{wildeboer}
 \bibinfo{author}{\bibfnamefont{J.~Wildeboer}} \bibnamefont{and}
 \bibinfo{author}{\bibfnamefont{A.~Seidel}},
 \bibinfo{title}{\bibfnamefont{Correlation Functions in $SU(2)$ Invariant Resonating-Valence-Bond Spin Liquids on Nonbipartite Lattices}},
  \bibinfo{journal}{Phys.~Rev.~Lett.} \textbf{\bibinfo{volume}{109}},
  \bibinfo{pages}{147208} (\bibinfo{year}{2012}).

\bibitem{yangfan}
 \bibinfo{author}{\bibfnamefont{F.~Yang}} \bibnamefont{and}
 \bibinfo{author}{\bibfnamefont{H.~Yao}},
 \bibinfo{title}{\bibfnamefont{Frustrated Resonating Valence Bond States in Two Dimensions: Classification and Short-Range Correlations}},
  \bibinfo{journal}{Phys.~Rev.~Lett} \textbf{\bibinfo{volume}{109}},
  \bibinfo{pages}{147209} (\bibinfo{year}{2012}).

%\bibitem{jiang}
%  \bibinfo{author}{\bibfnamefont{H.-C.~Jiang}},
 % \bibinfo{author}{\bibfnamefont{H.~Yao}} \bibfnamefont{and}
 % \bibinfo{author}{\bibfnamefont{L.~Balents}},
  %\bibinfo{title}{\bibfnamefont{Spin liquid ground state of the spin-1/2 square $J_1-J_2$ Heisenberg model}},
 % \bibinfo{journal}{Phys.~Rev.~B} \textbf{\bibinfo{volume}{86}},
  %\bibinfo{pages}{024424} (\bibinfo{year}{2012}).

\bibitem{anders_j1j2}
  \bibinfo{author}{\bibfnamefont{A.~Sandvik}},
  \bibinfo{title}{\bibfnamefont{Finite-size scaling and boundary effects in two-dimensional valence-bond-solids}},
  \bibinfo{journal}{Phys.~Rev.~B} \textbf{\bibinfo{volume}{85}},
  \bibinfo{pages}{134407} (\bibinfo{year}{2012}).

\bibitem{RVB}
  \bibinfo{author}{\bibfnamefont{P.~W.}~\bibnamefont{Anderson}},
  \bibinfo{title}{\bibfnamefont{The resonating valence bond state in La$_2$CuO$_4$ and superconductivity}},
  \bibinfo{journal}{Science} \textbf{\bibinfo{volume}{235}},
  \bibinfo{pages}{1196} (\bibinfo{year}{1987}).

\bibitem{patrick}
  \bibinfo{author}{\bibfnamefont{P.~A.~Lee}},
  \bibinfo{author}{\bibfnamefont{Naoto Nagaosa}} \bibnamefont{and}
  \bibinfo{author}{\bibfnamefont{X.-G.~Wen}},
  \bibinfo{title}{\bibfnamefont{Doping a Mott Insulator: Physics of High Temperature Superconductivity}},
  \bibinfo{journal}{Rev. Mod. Phys.} \textbf{\bibinfo{volume}{78}},
  \bibinfo{pages}{17} (\bibinfo{year}{2006}).

\bibitem{neutrons}
  \bibinfo{author}{\bibfnamefont{J.~M.~Tranquada}},
  \bibinfo{author}{\bibfnamefont{B.~J.~Sternlieb, J.~D.~Axe, Y.~Nakamura}} \bibnamefont{and}
  \bibinfo{author}{\bibfnamefont{S.~Uchida}},
  \bibinfo{title}{\bibfnamefont{Evidence for stripe correlations of spins and holes in copper oxide superconductors}},
  \bibinfo{journal}{Nature} \textbf{\bibinfo{volume}{375}},
  \bibinfo{pages}{561} (\bibinfo{year}{1995}).

\bibitem{stm}
  \bibinfo{author}{\bibfnamefont{
    M. J. Lawler,	
    K. Fujita,	
    Jhinhwan Lee,	
    A. R. Schmidt,	
    Y. Kohsaka,	
    Chung Koo Kim,	
    H. Eisaki,	
    S. Uchida,	
    J. C. Davis,	
    J. P. Sethna	}}
  \bibnamefont{and}
  \bibinfo{author}{\bibfnamefont{Eun-Ah Kim}},
  \bibinfo{title}{\bibfnamefont{Intra-unit-cell electronic nematicity of the high-Tc copper-oxide pseudogap states}},
  \bibinfo{journal}{Nature} \textbf{\bibinfo{volume}{466}},
  \bibinfo{pages}{374} (\bibinfo{year}{2010}).

\bibitem{nmr}
  \bibinfo{author}{\bibfnamefont{Tao Wu,	
Hadrien Mayaffre,	
Steffen Kr\"amer,	
Mladen Horvatic,	
Claude Berthier,	
W. N. Hardy,	
Ruixing Liang,	
D. A. Bonn, M.-H.~Julien}} \bibnamefont{and}
  \bibinfo{author}{\bibfnamefont{Marc-Henri Julien}},
  \bibinfo{title}{\bibfnamefont{Magnetic-field-induced charge-stripe order in the high-temperature superconductor YBa$_2$Cu$_3$O$_y$}},
  \bibinfo{journal}{Nature} \textbf{\bibinfo{volume}{477}},
  \bibinfo{pages}{191} (\bibinfo{year}{2011}).

\bibitem{xray}
  \bibinfo{author}{\bibfnamefont{Peter Abbamonte, Eugene Demler, J. C. Seamus Davis}} \bibnamefont{and}
  \bibinfo{author}{\bibfnamefont{Juan-Carlos Campuzano}},
  \bibinfo{title}{\bibfnamefont{Resonant soft x-ray scattering, stripe order, and the electron spectral function in cuprates}},
  \bibinfo{journal}{Physica C, Special Issue 
  ``Stripes and Electronic Liquid Crystals in Strongly Correlated Systems.", arXiv:1112.5112}
  %,\bibinfo{pages}{} 
  (\bibinfo{year}{2012}).

\bibitem{qdm1}
 \bibinfo{author}{\bibfnamefont{D.~Poilblanc}},
  \bibinfo{title}{\bibfnamefont{Properties of holons in the Quantum Dimer Model}},
  \bibinfo{journal}{Phys. Rev. Lett.} \textbf{\bibinfo{volume}{100}},
  \bibinfo{pages}{157206} (\bibinfo{year}{2009}).

\bibitem{qdm2}
  \bibinfo{author}{\bibfnamefont{C.~A.~Lamas}},
  \bibinfo{author}{\bibfnamefont{A.~Ralko}},
  \bibinfo{author}{\bibfnamefont{M.~Oshikawa}},
  \bibinfo{author}{\bibfnamefont{D.~Poilblanc}} \bibnamefont{and}
  \bibinfo{author}{\bibfnamefont{P.~Pujol}},
  \bibinfo{title}{\bibfnamefont{Statistics of holes and nature of superfluid phases in Quantum dimer models}},
  \bibinfo{journal}{Phys.~Rev.~B} \textbf{\bibinfo{volume}{87}},
  \bibinfo{pages}{104512} (\bibinfo{year}{2013}).

\bibitem{note1}
To guaranty full translational invariance, singlets are oriented from left to right, bottom to top. At $x=0$, this is equivalent 
to orient them from one sublattice to the other, up to the insertion of a single vison line in the cylinder ~\cite{didier,wang}.

\bibitem{corboz}
  \bibinfo{author}{\bibfnamefont{Philippe Corboz}},
  \bibinfo{author}{\bibfnamefont{Roman Orus}},
  \bibinfo{author}{\bibfnamefont{Bela Bauer}}  \bibfnamefont{and}
  \bibinfo{author}{\bibfnamefont{Guifre Vidal}},
  \bibinfo{title}{\bibfnamefont{Simulation of strongly correlated fermions in two spatial dimensions with fermionic Projected Entangled-Pair States}},
  \bibinfo{journal}{Phys.~Rev.~B} \textbf{\bibinfo{volume}{81}},
  \bibinfo{pages}{165104} (\bibinfo{year}{2010}).

\bibitem{kotliar}
  \bibinfo{author}{\bibfnamefont{G.~Kotliar}},
  \bibinfo{title}{\bibfnamefont{Resonating valence bonds and d-wave superconductivity}},
  \bibinfo{journal}{Phys.~Rev.~B} \textbf{\bibinfo{volume}{37}},
  \bibinfo{pages}{3664} (\bibinfo{year}{1988}).

\bibitem{supplement}
   \bibinfo{title}{\bibfnamefont{See Supplemental Material at http://xxx which includes
   results on the density-density correlations in the RVB superconductors and details on the iPEPS calculations.}}

 \bibitem{corboz2}
  \bibinfo{author}{\bibfnamefont{Philippe Corboz, T.~M.~Rice}} \bibfnamefont{and}
  \bibinfo{author}{\bibfnamefont{Matthias~Troyer}},
  \bibinfo{title}{\bibfnamefont{Competing states in the t-J model: uniform d-wave state versus stripe state}},
  \bibinfo{pages}{arXiv:14022859} (\bibinfo{year}{2014}).

\bibitem{coldea}
  \bibinfo{author}{\bibfnamefont{R. Coldea, S.M. Hayden, G. Aeppli, T.G. Perring, C.D. Frost, T.E. Mason, S.-W. Cheong}}  \bibfnamefont{and}
  \bibinfo{author}{\bibfnamefont{Z.~Fisk}},
  \bibinfo{title}{\bibfnamefont{Spin waves and electronic interactions in La$_2$CuO$_4$}},
  \bibinfo{journal}{Phys.~Rev.~Lett.} \textbf{\bibinfo{volume}{86}},
  \bibinfo{pages}{5377} (\bibinfo{year}{2001}).

\bibitem{malrieu}
  \bibinfo{author}{\bibfnamefont{Carmen J. Calzado}}  \bibfnamefont{and}
  \bibinfo{author}{\bibfnamefont{Jean-Paul Malrieu}},
  \bibinfo{title}{\bibfnamefont{Proposal of an extended t--J Hamiltonian for high-T$_c$ cuprates from ab initio calculations
on embedded clusters}},
  \bibinfo{journal}{Phys.~Rev.~B} \textbf{\bibinfo{volume}{63}},
  \bibinfo{pages}{214520} (\bibinfo{year}{2001}).

\bibitem{yao}
  \bibinfo{author}{\bibfnamefont{Dao-Xin Yao}}  \bibfnamefont{and}
  \bibinfo{author}{\bibfnamefont{E.~W.~Carlson}},
  \bibinfo{title}{\bibfnamefont{Magnetic excitations in the high-T$_c$ iron pnictides}},
  \bibinfo{journal}{Phys.~Rev.~B} \textbf{\bibinfo{volume}{78}},
  \bibinfo{pages}{052507} (\bibinfo{year}{2008}).

%\bibitem{abrahams}
 % \bibinfo{author}{\bibfnamefont{Qimiao Si}}  \bibfnamefont{and}
  %\bibinfo{author}{\bibfnamefont{Elihu Abrahams}},
  %\bibinfo{title}{\bibfnamefont{Strong Correlations and Magnetic Frustration in the High T$_c$ Iron Pnictides}},
  %\bibinfo{journal}{Phys.~Rev.~B} \textbf{\bibinfo{volume}{78}},
  %\bibinfo{pages}{052507} (\bibinfo{year}{2008}).

\bibitem{abrahams}
  \bibinfo{author}{\bibfnamefont{Pallab Goswami, Rong Yu, Qimiao Si}}  \bibfnamefont{and}
  \bibinfo{author}{\bibfnamefont{Elihu Abrahams}},
  \bibinfo{title}{\bibfnamefont{Spin dynamics of a J$_1$-- J$_2$ antiferromagnet and its implications for iron pnictides}},
  \bibinfo{journal}{Phys.~Rev.~B} \textbf{\bibinfo{volume}{84}},
  \bibinfo{pages}{155108} (\bibinfo{year}{2011}).

%\bibitem{kapitulnik}
  %\bibinfo{author}{\bibfnamefont{Jing Xia, Elizabeth Schemm, G. Deutscher, 
  %S. A. Kivelson, D. A. Bonn, W. N. Hardy, R. Liang, W. Siemons, G. Koster, M. M. Fejer}}  \bibfnamefont{and}
  %\bibinfo{author}{\bibfnamefont{A.~Kapitulnik}},
  %\bibinfo{title}{\bibfnamefont{Polar Kerr-Effect Measurements of the High-Temperature YBa$_2$Cu$_3$O$_{6+x}$ 
 % Superconductor: Evidence for Broken Symmetry near the Pseudogap Temperature}},
  %\bibinfo{journal}{Phys.~Rev.~Lett.} \textbf{\bibinfo{volume}{100}},
  %\bibinfo{pages}{127002} (\bibinfo{year}{2008}).

\bibitem{campuzano}
  \bibinfo{author}{\bibfnamefont{A. Kaminski, S. Rosenkranz, H. M. Fretwell, J. C. Campuzano,
Z. Li, H. Raffy, W. G. Cullen, H. You, C. G. Olson, C. M. Varma}}  \bibfnamefont{and}
  \bibinfo{author}{\bibfnamefont{H. H\"ochst}},
  \bibinfo{title}{\bibfnamefont{Spontaneous breaking of time reversal
symmetry in the pseudogap state of a high-T$_c$ superconductor}},
  \bibinfo{journal}{Nature} \textbf{\bibinfo{volume}{416}},
  \bibinfo{pages}{610} (\bibinfo{year}{2002}).

\bibitem{kapitulnik}
  \bibinfo{author}{\bibfnamefont{Hovnatan Karapetyan, Jing Xia, M. Hucker, G. D. Gu, J. M. Tranquada, M.M. Fejer}}  
  \bibfnamefont{and}
  \bibinfo{author}{\bibfnamefont{A.~Kapitulnik}},
  \bibinfo{title}{\bibfnamefont{Evidence of Chiral Order in the Charge-Ordered Phase of Superconducting 
  La$_{1.875}$Ba$_{0.125}$CuO$_4$ Single Crystals Using Polar Kerr-Effect Measurements}},
  \bibinfo{journal}{Phys.~Rev.~Lett.} \textbf{\bibinfo{volume}{112}},
  \bibinfo{pages}{047003} (\bibinfo{year}{2014}).

\bibitem{fauque}
  \bibinfo{author}{\bibfnamefont{B. Fauqu\'e, Y. Sidis, V. Hinkov, S. Pailh\`es, C. T. Lin, X.~Chaud}}  \bibfnamefont{and}
  \bibinfo{author}{\bibfnamefont{P.~Bourges}},
  \bibinfo{title}{\bibfnamefont{Magnetic Order in the Pseudogap Phase of High-T$_c$ Superconductors}},
  \bibinfo{journal}{Phys.~Rev.~Lett.} \textbf{\bibinfo{volume}{96}},
  \bibinfo{pages}{197001} (\bibinfo{year}{2006}).



\end{thebibliography}
\end{document}

\title{RVB superconductors with fermionic projected entangled pair states: \textit{supplemental material}}

\author{Didier Poilblanc}
\affiliation{Laboratoire de Physique Th\'eorique, C.N.R.S. and Universit\'e de Toulouse, 31062 Toulouse, France}

\author{Philippe Corboz}
\affiliation{Theoretische Physik, ETH-Zurich, 8093 Zurich, Switzerland}
\affiliation{Institute for Theoretical Physics, University of Amsterdam, Science Park 904 Postbus 94485, 1090 GL Amsterdam, The Netherlands}

\author{Norbert \surname{Schuch}}
\affiliation{Institut f\"ur Quanteninformation, RWTH Aachen, D-52056 Aachen, Germany}

\author{J.~Ignacio \surname{Cirac}}
\affiliation{Max-Planck-Institut f{\"{u}}r Quantenoptik,
Hans-Kopfermann-Str.\ 1, D-85748 Garching, Germany}

\date{\today}

\maketitle

\section{Density-density correlations and Coulomb energy}

\begin{figure}[htb]
\begin{center}
\includegraphics[width=10cm,angle=0]{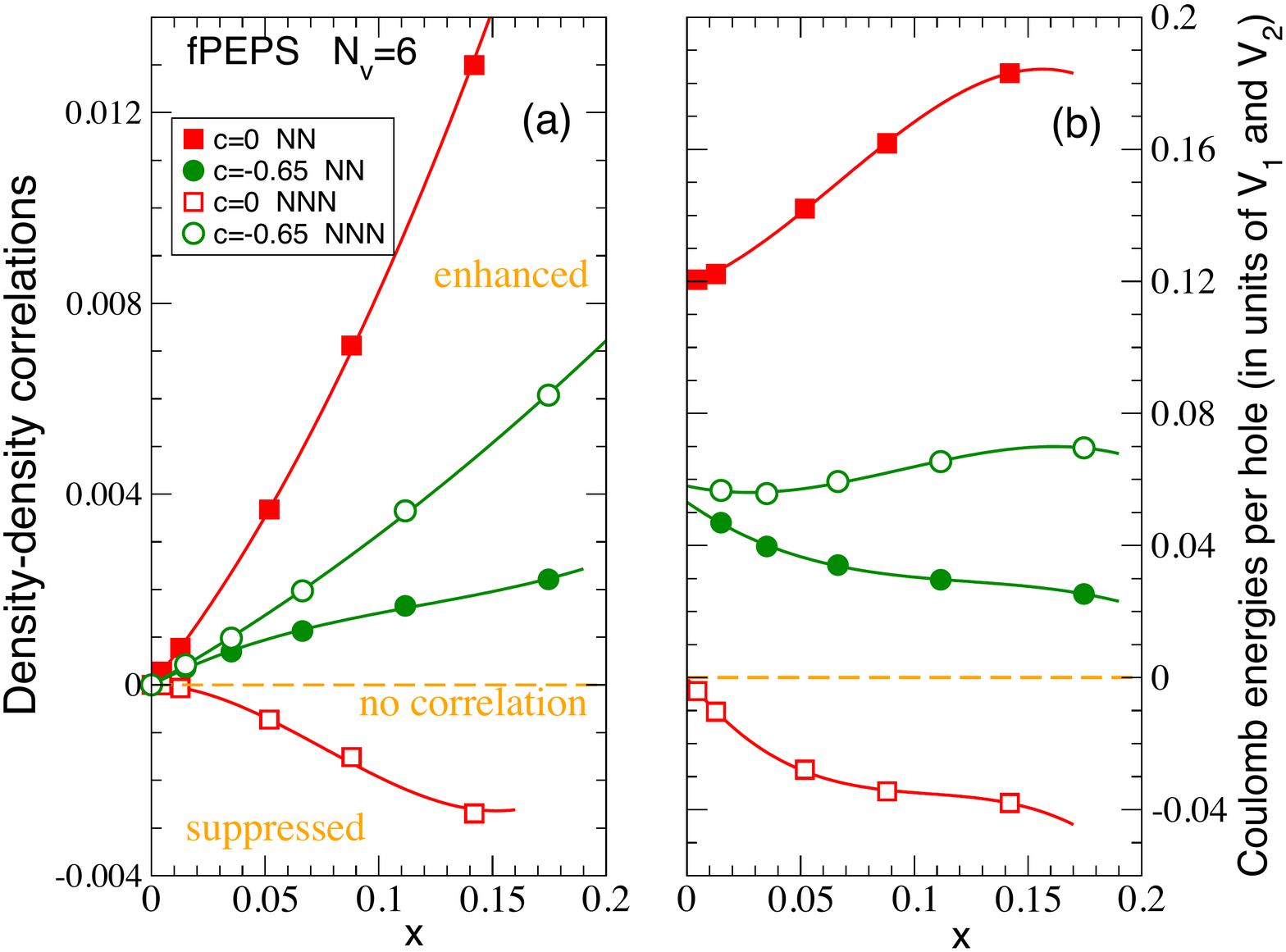}
\caption{(a) Density-density correlations between NN and NNN sites as a function of doping. The two RVB $d+is$ superconductors with
$c=0$ (pure $d+is$ RVB) and $c=-0.65$ (optimized RVB) are compared (computations made on an infinitely long $N_v=6$ cylinder). 
(b) NN and NNN Coulomb energies per hole. }
\label{Fig:coulomb}
\end{center}
\end{figure}

Superconducting (Cooper) pairs are bound states of charged electrons (or holes) and, hence, are expected to be 
destabilized by strong short range (SR) Coulomb repulsion. In cuprates superconductors the NN repulsion $V_1$ between doped holes is expected 
to be large, typically $V_1\sim t$ or larger~\cite{malrieu2001}, and the superconducting coherence length is short. It is therefore interesting to investigate the related energy cost in a RVB superconductor.
One can take into account the SR Coulomb repulsion by adding to the (extended) t--J model a term like:
\begin{equation}
H_{\rm V}=V_1\sum_{\big<ij\big>} ({n}_i-x)({n}_j-x) + V_2\sum_{\big<\big<kl\big>\big>} ({n_k}-x)({n}_l-x) +\cdots
\label{Eq:HV}
\end{equation}
where $n_i=1-c_{i\sigma}^\dagger c_{i\sigma}$ is the local hole density operator and the repulsion beyond NN ($V_1>0$) and NNN ($V_2>0$) sites can 
safely be neglected. Note that we have subtracted the energy cost $2V_1x^2+2V_2x^2$ (per site) of an uncorrelated uniform charge background.
The variational energies of both terms are therefore proportional to the corresponding {\it connected} density-density correlators,
\begin{equation}
C_{ij}=\big< ({n}_i-x)({n}_j-x)\big>  .
\end{equation}
The NN and NNN correlators $C_{NN}$ and $C_{NNN}$ are shown in Fig.~\ref{Fig:coulomb}(a) for two different RVB ans\"atze, (i) the $c=0$ pure $d+is$ superconductor (with only NN pairing)
and (ii) the optimized $c=-0.65$ (predominantly) $d$-wave superconductor. $C_{ij}>0$ ($C_{ij}<0$) corresponds to an enhanced (suppressed)
density correlation w.r.t. a uniform background of uncorrelated holes.
The strong NN correlation in the pure $s+id$ RVB superconductor is consistent with NN hole (Cooper) pairs and a superconducting coherence length
of just one lattice spacing. For the optimized RVB the largest density correlation occurs between NNN sites suggesting that Cooper pairs are  
predominantly NNN hole pairs despite the dominant $d$-wave character of the pairing field, in agreement with early exact diagonalisations  on small t--J clusters~\cite{poilblanc1994}.

Short distance pairing yields an increase of Coulomb energy $2V_1 C_{NN}/x + 2V_2 C_{NNN}/x$  per hole.  
Each term is plotted separately in Fig.~\ref{Fig:coulomb}(b) for the two RVB wave functions studied here.
Assuming $V_2\ll V_1$, we observe that the energy cost is very much reduced in the optimized RVB compared to the reference $c=0$ state, due to
a larger coherence length (or Cooper pair size).
Quantitatively, one finds a Coulomb energy of $\sim 0.04 V_1$ (per hole) which remains quite small compared to the
{\it gain} of kinetic energy $|e_k|\sim 2.5 t$ (per hole), even if $V_1>t$.  

\section{Details about the iPEPS calculations}

The (optimized) $d+is$ RVB superconductor has been compared to fermionic iPEPS calculations using the same bond dimension $D=3$ for the same Hamiltonian parameters i.e. $J_2=0.5 J_1$ and $J_1/t=0.4$ as in the paper (see Ref.~\cite{corboz2010} for an introduction to the method). We used two different optimization methods, the simple update  from Ref.~\cite{corboz2010-nn} and the full update method from Ref.~\cite{Corboz13_shastry} for Hamiltonians with next-nearest neighbor interactions. The former approach is computationally cheaper but less accurate than the latter one (see Ref.~\cite{corboz2010} for a discussion). Both optimization methods require a $2\times 2$ unit cell of tensors (or larger) which is periodically repeated on the lattice. 

\begin{figure}[htb]
\begin{center}
\includegraphics[width=10cm,angle=0]{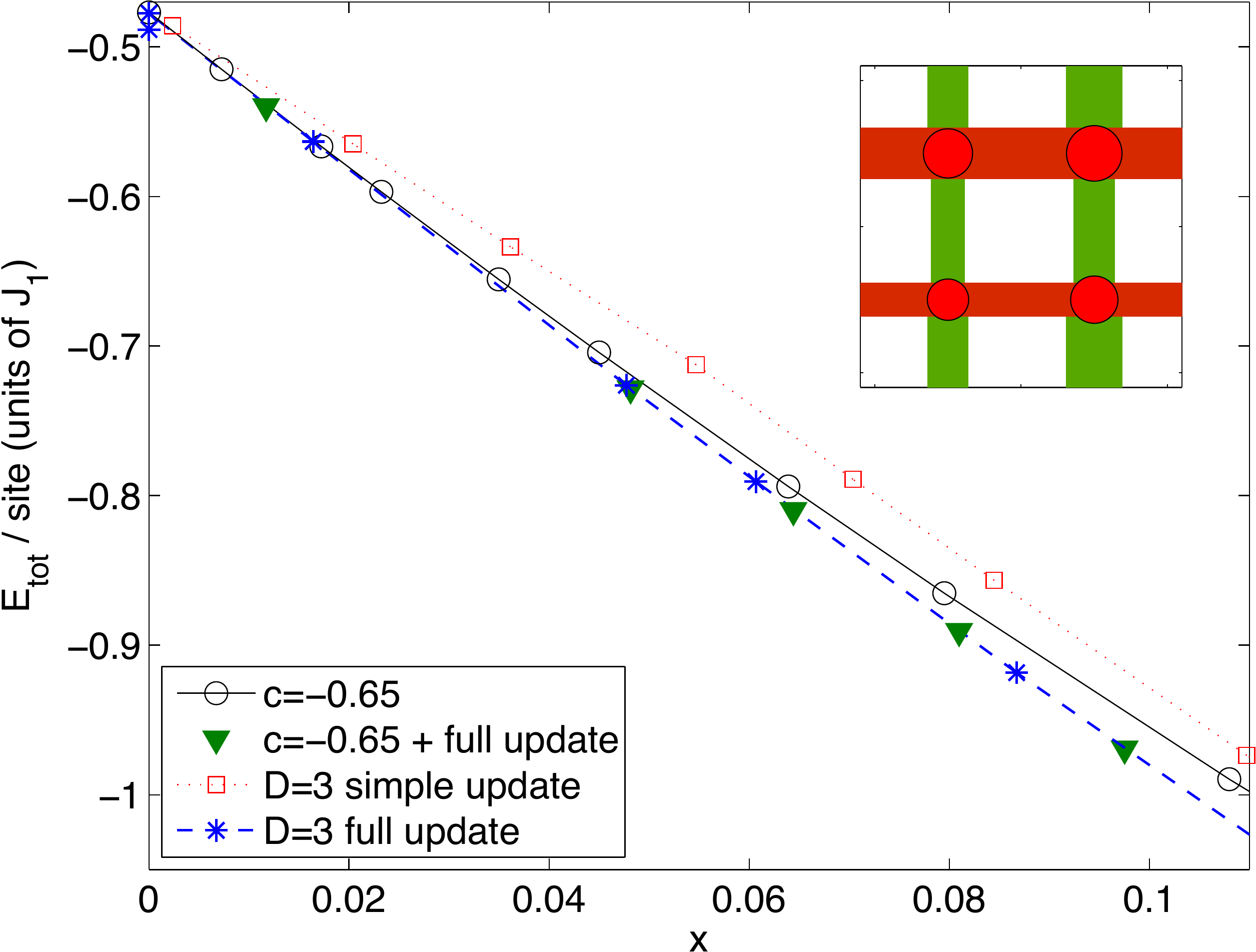}
\caption{Total energy (per site) of the d+is ansatz with $c=-0.65$ and the optimized $D=3$ iPEPS wavefunctions (using different update methods and initial states) as a function of doping.   Inset: Real space pattern of the best $D=3$ iPEPS state within the $2\times 2$ unit cell for $x=0.091$. The red dots show the hole density, and the colored bonds the pairing amplitude with different signs.}
\label{Fig:ipeps_ener}
\end{center}
\end{figure}

Figure~\ref{Fig:ipeps_ener} shows that the best state obtained with the full update optimization with $D=3$ (blue stars) is only slightly lower than the constructed $c=-0.65$ RVB state (open black circles), which is remarkable given the fact that the latter has only one single variational parameter. These results have been obtained starting from a random initial state (with the same random tensor on each lattice site), and a similar result is obtained by using the $c=-0.65$ RVB ansatz as initial state for the full optimization (green triangles). We note that depending on the random initial state some of the simulations also converged to other (competing) states, but all having a higher energy. The simple update method (red squares) fails to find the best state in this case, but yields states with a higher energy. All these results show that the (kinetic energy) optimized $d+is$ RVB state is an excellent $D=3$ ansatz for this range of Hamiltonian parameters, and it can serve as an ideal initial state for fermionic iPEPS to study RVB phases, e.g. to perform simulations at larger~$D$.

We note that there is a jump in the energy per site in the iPEPS data at $x=0$. The reason is that for $x=0$ there are no charge degrees of freedom and the iPEPS can fully exploits its auxiliary space of dimension $D=3$ for the spin degrees of freedom, whereas for any finite $x$ the iPEPS needs to carry also information on the charge degrees of freedom through its auxiliary space. In other words, since for $x=0$ there are less degrees of freedom the $D=3$ iPEPS lies closer to the true ground state than the one at finite $x$, which leads to a discontinuity in the energy per site. For this reason we computed the energy per hole in the main text as $E_{hole}=\left( \langle H \rangle_x - \langle H \rangle_{0+\epsilon} \right)/x$ where we take $\epsilon \rightarrow 0$. Finally we note that also the energy of the RVB ansatz for $x=0$ can also be further improved, see Ref.~\cite{wang13}.

An important feature of the RVB $d+is$ superconductor is that it is translationally invariant (i.e. it is uniform)
and its bond SC amplitudes are invariant under all (rotation and/or reflection) symmetry operations of the 
lattice $C_{4v}$ point group (although minor differences between vertical and horizontal directions occur
on finite circumference cylinders due to small finite-size effects). Furthermore it preserves the SU(2) spin rotation symmetry. The best optimized iPEPS reproduces these symmetries approximately, in particular the magnetic moment at each site vanishes entirely (i.e. the SU(2) symmetry is unbroken), and there are  only little variations in the hole density per site and the bond SC amplitudes (see inset of  Fig.~\ref{Fig:ipeps_ener}). This slight symmetry breaking can be partly due to the Trotter error in the optimization method and/or to the fact that a finite bond dimension can also artificially induce a symmetry breaking (which typically gets fully restored for large $D$).

Using larger unit cells we expect to get stripe states similarly as in the plain t--J model~\cite{corboz14} competing with
the uniform RVB superconductor. A full study using the full update is challenging (more challenging than in the plain t--J model 
because of the next-nearest neighbor interaction which leads to a higher computational cost) and is left for further studies.

\bibliographystyle{apsrev4-1}
\bibliography{refs}